\documentclass[useAMS,usenatbib]{mn2e}

\usepackage{amssymb}
\usepackage{aas_macros}
\usepackage{natbib}
\usepackage{times}
\usepackage{graphicx}

\newcommand{\Myr}                      {\,{\rm Myr}}

\newcommand{\Gyr}                      {\,{\rm Gyr}}

\newcommand{\Msun}                    {\,{\rm M}_\odot}
\newcommand{\Msunyr}                 {\,{\rm M}_\odot\,{\rm yr}^{-1}}

\newcommand{\erg}                       {\,{\rm erg}}

\newcommand{\K}                          {\,{\rm K}}
\newcommand{\keV}                      {\,{\rm keV}}

\newcommand{\gimic}        {\textsc{gimic}}

\newcommand{\apec}              {\textsc{Apec}}

\newcommand{\einstein}             {\textit{Einstein}}
\newcommand{\rosat}             {\textit{ROSAT}}
\newcommand{\chandra}         {\textit{Chandra}}
\newcommand{\xmm}             {\textit{XMM-Newton}}
\newcommand{\ixo}             {\textit{IXO}}
\newcommand{\iras}             {\textit{IRAS}}
\newcommand{\galex}             {\textit{GALEX}}
\newcommand{\twomass}       {\textsc{2Mass}}
\newcommand{\ipac}              {\textsc{Ipac}}

\newcommand{\sfr}                {$\dot{M}_\star$}
\newcommand{\lx}                 {$L_{\rm X}$}
\newcommand{\lk}                 {$L_{\rm K}$}

\newcommand{\mvir}              {$M_{\rm vir}$}
\newcommand{\lxlk}              {$L_{\rm X} - L_{\rm K}$}

\newcommand{\lxsfr}             {$L_{\rm X} - \dot{M}_\star$}
\newcommand{\lxTx}              {$L_{\rm X} - T_{\rm X}$}

\def\lsim{\mathrel{\lower0.6ex\hbox{$\buildrel {\textstyle <}
 \over {\scriptstyle \sim}$}}}

\title[X-ray coronae of $L_\star$ galaxies]{The similarity of observed X-ray coronae associated with $L_\star$ disc and elliptical galaxies}
 \author[R.~A.~Crain et al.]  {\parbox[h]{160mm} { 
    Robert A. Crain$^{1}$\thanks{E-mail: rcrain@astro.swin.edu.au}, 
    Ian G. McCarthy$^{2}$,
    Joop Schaye$^{3}$,
    Carlos S. Frenk$^{4}$ \\
    and Tom Theuns$^{4,5}$}
  \vspace{6pt}\\
  $^1$Centre for Astrophysics \& Supercomputing, Swinburne University of Technology, Hawthorn, Victoria 3122, Australia\\
  $^2$Kavli Institute for Cosmology, University of Cambridge, Madingley Road, Cambridge, CB3 0HA \\
  $^3$Leiden Observatory, Leiden University, PO Box 9513, 2300 RA Leiden, Netherlands \\
  $^4$Institute for Computational Cosmology, Department of
  Physics, University of Durham, South Road, Durham, DH1 3LE\\
  $^5$Department of Physics, University of Antwerp, Campus Groenenborger, Groenenborgerlaan 171, B-2020 Antwerp,  Belgium}

\begin{document}

\date{\today}
\pagerange{\pageref{firstpage}--\pageref{lastpage}} \pubyear{2009}

\maketitle

\label{firstpage}

\begin{abstract}
The existence of hot, X-ray luminous gaseous coronae surrounding
present day $L_\star$ galaxies is a generic prediction of galaxy
formation theory in the cold dark matter cosmogony.  While extended
X-ray emission has been known to exist around elliptical galaxies for
 a long time, diffuse extra-planar emission has only recently been
detected around disc galaxies. We compile samples of elliptical and
disc galaxies that have \chandra\ and \xmm\ measurements, and compare
the scaling of the coronal X-ray luminosity ($L_{\rm X}$) with both the $K$-band
luminosity ($L_{\rm K}$) and the coronal X-ray temperature ($T_{\rm
X}$). The X-ray flux measurements 
are corrected for non-thermal point source contamination by spatial
excision and spectral subtraction for resolved and unresolved sources
respectively.  We find that the properties of the extended X-ray
emission from galaxies of different morphological types are similar:
for both elliptical and disc galaxies, the \lxlk\ and \lxTx\ relations
have similar slope, normalisation and scatter. The observed
universality of coronal X-ray properties suggests that the bulk of
this emission originates from gas that has been accreted, shock-heated
and compressed during the assembly of the galaxy and that outflows
triggered by stellar processes make only a minor contribution to the
X-ray emission.  This reservoir of cooling gas is a potential source
of morphological transformation; it provides a fresh supply of
material for discs to grow around galaxies of all morphological
types. 
\end{abstract}


\section{Introduction}
\label{sec:introduction}

In the cold dark matter (CDM) cosmogony, galaxy formation is a
continuous process fuelled by the accretion of material in the form of
dark matter, stars and gas.  Building upon the general idea of 
hierarchical clustering proposed by
\citet{White_and_Rees_78}, \citet[][hereafter
WF91]{White_and_Frenk_91} developed an analytical framework to
calculate galaxy formation in a CDM universe under certain simplifying
assumptions, such as spherical symmetry and isothermal density
profiles.  In this model, when sufficiently massive dark matter halos
collapse, their associated gas is shock-heated to the virial
temperature of the halo, forming a hot, quasi-hydrostatic corona from which
gas slowly cools through line emission and thermal bremsstrahlung,
feeding a disc around the growing central galaxy. A central prediction
of this model is that the cooling radiation from present-day $L_\star$
galaxies should escape as soft ($k_{\rm B}T_{\rm X} \sim 0.1\keV$)
X-rays, and thus that these galaxies should be surrounded by extended
X-ray coronae.

Although in the WF91 model there is no distinction between the nature
of the coronae surrounding disc or elliptical galaxies, isolated
massive spiral galaxies were identified by
\citet{Benson_et_al_00} as particularly promising targets for
detecting X-ray coronae. However, the {\em Einstein} and
\rosat\ observatories failed to provide any evidence for this, 
setting instead upper limits
\citep[e.g.][]{Bregman_and_Glassgold_82,Vogler_Pietsch_and_Kahabka_95,Bregman_and_Houck_97,Fabbiano_and_Juda_97,Benson_et_al_00}.
The first detections of circumgalactic X-ray emission around local
disc galaxies were finally possible with the \chandra\ and \xmm\
telescopes but this emission turned out to be one to two orders of
magnitude fainter than predicted by WF91
\citep{Strickland_et_al_04,Wang_05,Tullmann_et_al_06,Li_Wang_and_Hameed_07,Jeltema_Binder_and_Mulchaey_08,Owen_and_Warwick_09,Rasmussen_et_al_09,Sun_et_al_09}.

It is widely believed that the most likely source of soft X-ray
emission in many of these disc galaxies is outflowing gas driven by
energy from Type~II supernovae (SNe), rather than a cooling inflow
from a hot corona. The evidence for this is the observed
correlation between star formation rate and coronal soft X-ray
luminosity. In some particularly 
spectacular cases, such as M82, the X-ray emission exhibits a biconical
morphology suggestive of outflows 
\citep{Strickland_et_al_04}, and this could also be the case for  our
own Milky Way 
\citep{Bland-Hawthorn_and_Cohen_03,Su_Slatyer_and_Finkbeiner_10}.

Since the existence of hot gaseous coronae around sufficiently massive
galaxies is a quintessential component of the WF91 model (and thus of
the many semi-analytic models based on it), failure to detect the
predicted X-ray emission poses an interesting problem for the
canonical view of galaxy formation in a CDM universe\footnote{The
so-called ``cold flows'' advocated, for example, by
\cite{Birnboim_and_Dekel_03} are subdominant for the halo masses and
redshifts of interest here.}. A possible solution was recently proposed
by  
\citet[][hereafter C10]{Crain_et_al_10}. They showed that disc,
star-forming galaxies in the cosmological hydrodynamic simulations
of the \textsc{Galaxies-Intergalactic Medium Interaction Calculation}
\citep[\gimic,][]{Crain_et_al_09_short} do, in fact, develop the kind
of gaseous coronae predicted by WF91, but with an associated X-ray
emission that is one to two orders of magnitude fainter than the WF91
prediction. The main reason for this is that, contrary to the
assumption of WF91, the density profile of the hot gas does not follow
the density profile of the dark matter. Instead, it is much less
centrally concentrated as a result of energy injection from SNe at the
peak of the star formation activity, $z\sim1-3$. This raises the
entropy of the gas, both by driving outflows that shock the gas to
high temperatures and by ejecting low-entropy gas from the progenitor
haloes.

C10 showed that although the X-ray emission around disc galaxies in \gimic\ comes predominatly from an extended quasi-hydrostatic corona heated by shocks and gravitational contraction, the galaxies reproduce the observed scaling of the soft X-ray luminosity with $K$-band luminosity, disc rotation velocity and star formation rate. The presence of the latter correlation -- which arises because both the X-ray luminosity and the star formation rate scale with halo mass -- contradicts the main source of evidence in favour of the view that the X-ray emission observed around real disc galaxies must be directly associated with star formation (perhaps through a galactic wind or fountain). The low surface brightness of coronal gas renders it difficult to detect with \chandra\ and \xmm, making it an ideal target for future facilities such \textit{Astro-H} or the \textit{International X-Ray Observatory} (\ixo). Sensitive, high-resolution spectroscopy with these facilities has the potential to verify directly whether coronal gas is quasi-hydrostatic or outflowing.

In the meantime, it is interesting to explore another consequence of
the WF91 framework, namely that the properties of the hot coronal gas
should be broadly independent of the morphological characteristics of
the galaxy. Thus, just as bright disc galaxies, bright elliptical
galaxies are also expected to have extended coronae of X-ray emitting
hot gas.  Furthermore, since ellipticals are dominated by old stellar
populations and have little ongoing star formation, the view that most
of the X-ray emission comes from outflowing winds triggered by SNe-II
can be readily ruled out
\citep[for a review, see][]{Mathews_and_Brighenti_03}. Active galactic
nuclei (AGN) are not thought to be important in the production of
X-ray emission from $L_\star$ ellipticals \citep{David_et_al_06}.

In this paper we explore the properties of diffuse X-ray emission in
$L_\star$ galaxies of different optical morphologies.  Recent data
from \chandra\ and \xmm\ provide relatively large samples of X-ray
luminosities and temperatures (or upper limits) for normal (i.e.\
non-interacting, relatively isolated) $L_\star$ disc and elliptical
galaxies. The observational samples that we analyse in this paper are
introduced in \S\,\ref{sec:xray_data}. In
\S\,\ref{sec:scalings}, we present comparisons of derived X-ray scalings
for discs and ellipticals. In \S\,\ref{sec:interpretation}, we propose
that the hot coronal gas in disc and elliptical galaxies has a common
origin -- accretion onto a shock-heated quasi-hydrostatic hot corona -- and
explore the consequences of this proposal. We include a short
appendix, in which we show that ejecta from SNe-Ia and SNe-II are
unlikely to be the source of the hot coronal gas in galaxies of any
morphological type.


\section{X-ray data}
\label{sec:xray_data}

We first introduce the sample of galaxies that we use in this study,
which are taken from previously published X-ray studies. As the X-ray
flux from point sources can be comparable to (or even dominate) that
from the hot gas of normal galaxies, high quality X-ray observations
are required to determine the emission that truly originates from the
coronal gas. For this reason, we have limited our compilation of
galaxies to those that have \chandra\ and/or \xmm\ observations
reported in the literature. We describe the sample below, with a brief
discussion of how the X-ray luminosity of the hot gas was measured in
each case.

\subsection{Elliptical galaxies}
\label{sec:elliptical_galaxies}

Our sample of elliptical galaxies with X-ray luminosities and
temperatures is taken from two studies: \citet[][hereafter
D06]{David_et_al_06} and \citet[][hereafter
MJ10]{Mulchaey_and_Jeltema_10}. The D06 sample is comprised of 18 low
optical luminosity ($L_{\rm B} \la 3\times10^{10} L_{\rm B,\odot}$)
field elliptical galaxies observed with \chandra, drawn from a larger
sample of early type galaxies with \chandra\ archival data (C. Jones,
in prep.). The non-thermal X-ray emission from low-mass X-ray binaries
(LMXBs) was accounted for in D06 by spatial excision of bright sources
and spectral modelling (using a power-law component) of unresolved
sources. MJ10 selected their sample of 23 nearby field early type
galaxies from previously published X-ray catalogues
\citep[e.g.][]{O'Sullivan_Forbes_and_Ponman_01}. Approximately half of
their sample was observed with \chandra\ and the other half with \xmm,
with several galaxies having data from both satellites. The MJ10
sample nicely complements that of D06, as it is comprised of
relatively bright systems ($L_{\rm K} \ga 10^{11} L_{\rm
K,\odot}$). In similar fashion to D06, MJ10 accounted for non-thermal
emission from LMXBs through the inclusion of a power-law component in
their spectral modelling.

While it is possible to remove the contribution from unresolved
non-thermal sources on the basis of their spectra (which can be can be
differentiated from thermal spectra if there are sufficient photons),
removal of X-ray emission from a potential \textit{thermal} point
source population is more difficult. An good example is the emission
originating from the so-called Galactic Ridge. This emission had
previously been believed to come from hot gas, since its spectrum is
consistent with an optically-thin plasma (with $T\sim10^{7-8}$ K) and
even displays a prominent iron K line. However,
\citet{Revnivtsev_et_al_06,Revnivtsev_et_al_08} pointed out that the
X-ray surface brightness traces almost perfectly the $K$-band surface
brightness in that region of the Galaxy, as is also the case in
external galaxies such as NGC 3379 (which is part of the D06
sample). \citet{Revnivtsev_et_al_09} used a 1 Ms \chandra\ exposure to
show that, indeed, most of the Galactic Ridge emission originates from
individual faint point sources, specifically accreting white dwarfs
and cataclysmic variable stars.

At present it is not possible to remove directly the contribution of
faint thermal point sources to the X-ray luminosity in the samples of
D06 and MJ10. However, we can use the tight correspondence, reported
by \citet{Revnivtsev_et_al_08}, between the X-ray luminosity of these
faint sources and the $K$-band luminosity, to estimate their
importance. We do this below, in \S\,\ref{sec:LxLk}. We conclude from
this comparison that the contribution from faint thermal point sources
is potentially significant for 5 or 6 faint ellipticals from D06, and
for 3 ellipticals from MJ10.

\subsection{Disc galaxies}
\label{sec:disc_galaxies}

Unfortunately, there are no large, homogeneously analysed samples of
normal disc galaxies observed with \chandra\ or \xmm, analogous to
those of D06 and MJ10 for ellipticals. This may, in part, be due to
the commonly held belief (arising from \rosat's low detection rate of
coronal gas) that disc galaxies do not possess X-ray-luminous coronae,
suggesting that there is little point in obtaining X-ray observations
of such systems for the purpose of studying hot gas. In spite of this,
there is a growing body of work on small samples of disc galaxies that
shows that these galaxies do indeed have detectable diffuse, coronal
emission. As we will show below, the properties of the hot coronae of
disc galaxies are remarkably similar to those of ellipticals of the
same mass. 

Our heterogeneous sample of disc galaxies is taken from a number of
studies, including \citet[][hereafter Str04]{Strickland_et_al_04},
\citet[][hereafter W05]{Wang_05}, \citet[][hereafter
T06]{Tullmann_et_al_06}, \citet[][hereafter
L07]{Li_Wang_and_Hameed_07}, \citet[][hereafter
OW09]{Owen_and_Warwick_09}, \citet[][hereafter Sun07]{Sun_et_al_07},
\citet[][hereafter J08]{Jeltema_Binder_and_Mulchaey_08}, and
\citet[][hereafter R09]{Rasmussen_et_al_09}. 
Str04, W05, T06, L07, and R09 all studied \textit{edge-on} disc
galaxies. Str04 used \chandra\ to observe a sample of 10 star-forming
galaxies, 7 of which are classified as starbursts. W05 report on
\chandra\ observations of 7 `normal' star-forming galaxies. T06
observed with \xmm\ a sample of 9 normal star-forming disc
galaxies. L07 observed the nearly edge-on Sombrero galaxy (M104) with
\chandra. R09 observed two quiescent edge-on disc galaxies with
\chandra, but found no significant diffuse emission away from the
disc. For all these studies we use only the reported
\textit{extra-planar} X-ray luminosity (or upper limits) of the hot
gas (see Table~9 of Str04, Table~1 of W05, and Table~9 of T06).  

Since we exclude the luminosity from the region that is spatially
coincident with the disc, there is no significant contribution from
faint thermal point sources to the X-ray luminosities of disc galaxies
that we analyse here. OW09, however, observed a sample of 6 nearby
{\it face-on} disc galaxies with {\it XMM-Newton}. For these systems
emission from faint thermal point sources could be a contaminant but,
as we show below, the expected contribution from these sources is much
smaller than the total measured X-ray luminosities.

Finally, S07 and J08 reported {\it Chandra} X-ray detections (and
upper limits) of optically luminous early and late type galaxies in
several nearby galaxy groups and clusters. We have elected to use
their late type galaxies to complement our relatively small sample, in
spite of the fact that the luminosity of these systems could be
influenced by the group/cluster environment (e.g., if ram pressure
strips some of the hot gas). In a forthcoming study (McCarthy et al.\,
\textit{in prep}), however, we show (using cosmological hydrodynamical
simulations) that, for those galaxies that are not completely stripped
of their gas, the X-ray luminosity is largely unchanged by ram
pressure stripping. This is because the X-ray luminosity is very
centrally concentrated so the brightest gas is the very last to be
stripped.

\section{Comparison of the observed X-ray scalings of disc and
elliptical galaxies} 
\label{sec:scalings}

In this section, we examine correlations between the X-ray luminosity
of the hot gas and the near-infrared ($K$-band) luminosity of the galaxy,
as well as the temperature of the hot gas, inferred from X-ray
spectroscopy.

\subsection{The \lxlk\ relation of disc and elliptical galaxies}
\label{sec:LxLk}

\begin{figure}
\includegraphics[width=\columnwidth]{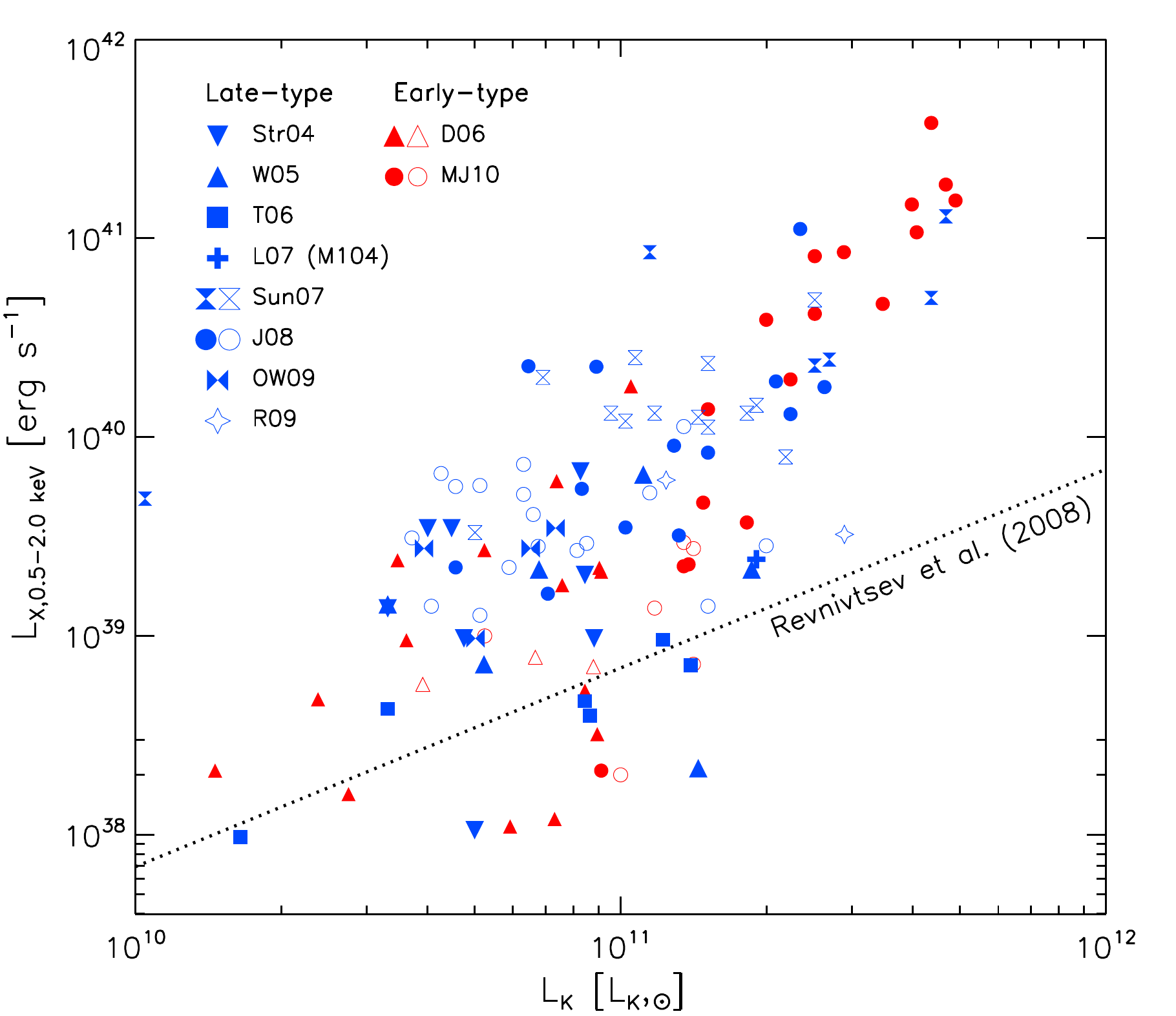}
\caption{The 0.5-2.0\keV\ X-ray luminosity of the hot gas as a
function of $K$-band luminosity, for a sample of disc (blue symbols)
and elliptical (red symbols) galaxies drawn from a number of studies
(see legend and text). For disc galaxies, we quote extra-planar X-ray
luminosities where possible and, for both morphological classes, X-ray
luminosities have been corrected for non-thermal point source
contributions. Filled symbols denote X-ray detections, open symbols
denote upper limits. Corresponding $K$-band luminosities were
extracted from the online \twomass\ database. The dotted line
represents the potential contribution from faint thermal point sources
(see text). Remarkably, the two morphological types exhibit broadly
the same slope, normalisation and scatter in the \lxlk\ relation.} 
\label{fig:LxLk} 
\end{figure}

We begin by examining, as a function of stellar morphology, the
scaling of the diffuse soft X-ray luminosity with the
$K$-band luminosity, \lk, which is a good proxy for stellar mass. We
have converted X-ray luminosities quoted in other passbands into the
0.5-2.0\keV\ band using the \textsc{pimms}
tool\footnote{http://heasarc.nasa.gov/docs/software/tools/pimms.html}
\citep{Mukai_93}. $K$-band luminosities were extracted from the
\textsc{ipac} online database for the Two Micron All-Sky Survey
\citep[\twomass,][]{Skrutskie_et_al_06_short}.  

Fig.~\ref{fig:LxLk} shows the coronal soft X-ray luminosity as a
function of $K$-band luminosity for our samples of disc (blue symbols)
and elliptical (red symbols) galaxies. Systems with X-ray detections
are denoted by filled symbols, whilst upper limits are denoted by open
symbols. Remarkably, the two morphological classes populate the
\lx-\lk\ plane in a very similar way: the relation between these
two properties has similar slope, normalisation and scatter for both
classes. We conclude that, for fixed stellar mass, the X-ray
luminosity of hot coronae is unrelated to the morphology of the host
galaxy.

Since the X-ray emission has been explicitly corrected for non-thermal
point-source contamination, the correlation in Fig.~\ref{fig:LxLk} is
not a reflection of the linear correlation between \textit{total}
X-ray luminosity (i.e. uncorrected for point sources) and optical
luminosity that is known to exist for low optical luminosity
ellipticals \citep{O'Sullivan_Forbes_and_Ponman_01}. Nor is the
correlation driven by a contribution from faint thermal point sources
(e.g. accreting white dwarfs and cataclysmic variable stars) that
cannot be removed spectrally, since only a small number of faint
ellipticals in our sample have coronal luminosities that are
comparable to, or less than, the integrated luminosity of thermal
point sources inferred from the relation of \citet[][see dotted line
in Fig.~\ref{fig:LxLk}]{Revnivtsev_et_al_08}. Several of our faint
disc galaxies also lie below this relation but, as discussed in
\S\,\ref{sec:disc_galaxies}, the luminosities from Str04, W05, T06,
L07, and R09 are attributed exclusively to \textit{extra-planar}
emission, and are therefore unlikely to be contaminated by point
sources.

The correlation between the optical and X-ray luminosities of disc and
elliptical galaxies has been explored previously
\citep[e.g.][]{Fabbiano_89}. However, such studies analysed data from
the \einstein\ and \rosat\ telescopes, which i) lacked the sensitivity
to detect diffuse X-ray emission in low (optical) luminosity galaxies
and ii) lacked the spatial and spectral resolution to enable the
subtraction of point-source contributions to the X-ray flux. As a
result, those studies were not able to find the similarity in the
correlation between the \textit{coronal} X-ray luminosity and stellar
mass for disc and elliptical galaxies that we have uncovered here.

\subsection{The \lxTx\ relation of disc and elliptical galaxies}
\label{sec:LxTx}

\begin{figure}
\includegraphics[width=\columnwidth]{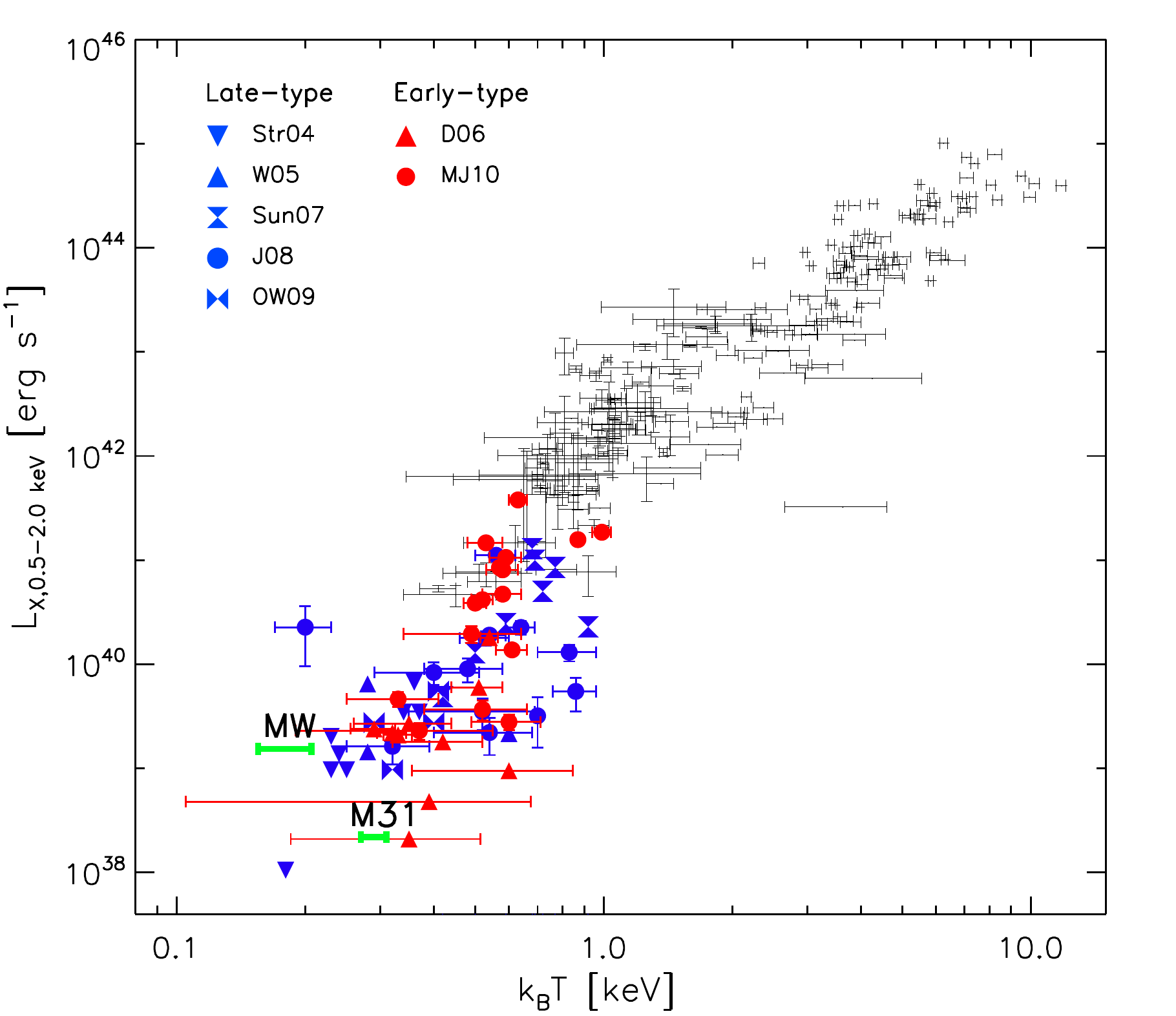}
\caption{The X-ray luminosity-temperature relation in the 0.5-2.0
keV band. We plot those galaxies from the sample shown in
Fig.~\ref{fig:LxLk} that i) have a spectroscopic measurement of the
coronal temperature and ii) in the case of ellipticals, have a total
X-ray luminosity above the expected thermal point source
contribution. Also plotted are measurements for the Milky Way
\citep{Henley_et_al_10} and M31 \citep{Liu_et_al_10}, shown as green
error bars, the galaxy group samples of
\citet{Helsdon_and_Ponman_00} and \citet{Mulchaey_et_al_03}, and the
galaxy cluster sample of \citet{Horner_01}, both shown as black error
bars.  Note the break in the \lxTx\ relation at $\sim1\keV$.  As was
the case for the \lxlk\ relation of Fig.~\ref{fig:LxLk}, disc and
elliptical galaxies follow the same relation.}
\label{fig:LxTx}
\end{figure}

The similarity of the \lxlk\ relations for disc and elliptical
galaxies revealed in Fig.~\ref{fig:LxLk} indicates that the X-ray
luminosity of hot coronal gas does not depend on the morphology of the
visible galaxy for systems of fixed mass, insofar as the $K$-band
luminosity reflects the stellar mass and the stellar mass reflects the
total mass. It is conceivable, however, that normal disc and
elliptical galaxies could have different stellar mass fractions and that the
similarity of their 
\lxlk\ relations could therefore be the result of some `conspiracy' or
coincidence. For example, ellipticals could be more X-ray luminous at
a fixed total mass, but also have higher stellar mass fractions. We
can rule out any potential conspiracy of this sort by examining the
\lxTx\ relation. The temperature of the gas is a measure of the depth
of the total (stars+gas+dark matter) potential well of the galaxy
\citep[e.g.][]{Voit_et_al_02}, so long as the gas is relatively close
to hydrostatic equilibrium. This is a reasonable assumption, since if
the gas were far from hydrostatic equilibrium, it would quickly
collapse or leave the system.

Fig.~\ref{fig:LxTx} shows the X-ray luminosity as a function of the
hot gas spectral temperature for those galaxies from the sample
presented in Fig.~\ref{fig:LxLk} that have temperature
estimates. Note, however, that we have excluded those galaxies from
the samples of D06 and MJ10 for which the inferred X-ray luminosity
lies below the estimated contribution from faint thermal point sources
(see discussion in \S~\ref{sec:elliptical_galaxies} and
\ref{sec:disc_galaxies}). For reference, we also include measurements
of galaxy groups, taken from the studies of
\citet{Helsdon_and_Ponman_00} and \citet{Mulchaey_et_al_03}, galaxy
clusters from Horner (2001), and of the Milky Way
\citep{Henley_et_al_10} and M31 \citep{Liu_et_al_10}. 

We find, once again, the remarkable result that disc and elliptical
galaxies follow the same relation. This provides a strong argument
against the notion that an astrophysical coincidence or conspiracy is
responsible for the similarity of the \lxlk\ relations for the two
morphological types. The relation shown in Fig.~\ref{fig:LxTx}
reinforces our previous conclusion that the X-ray properties of hot
coronal gas do not depend on stellar morphology. It is also
interesting to note that the addition of our galaxy samples to the
well-known \lxTx\ relation obeyed by galaxy groups and galaxy clusters
forms a broken power-law with the break at approximately $1\keV$. We
discuss this intriguing result further in
\S\,\ref{sec:interpretation}.  

\section{Interpretation and discussion}
\label{sec:interpretation}

The presence of hot, X-ray luminous coronae around present day
$L_\star$ galaxies is a fundamental prediction of galaxy formation
theory in a cold dark matter cosmology. Indeed, such hot coronae arise
in both analytic and numerical models
\cite[e.g. WF91, C10, see
also][]{Benson_et_al_00,Toft_et_al_02,Rasmussen_et_al_09}. They form
as gas accreting onto growing dark matter halos is shock-heated at the
virial radius and adiabatically compressed. In sufficiently large
halos, the cooling time is longer than the infall time and the gas
forms a quasi-hydrostatic atmosphere around the galaxy\footnote{When the
cooling time is short, galaxies can accrete gas without it being
shock-heated at the virial radius. However, C10 showed, using the
\gimic\ simulations, that these ``cold flows,''
\citep[e.g.][]{Birnboim_and_Dekel_03,Keres_et_al_05} provide only a
small fraction of the ongoing gas accretion onto $L_\star$ galaxies
today.}. As it slowly cools, radiating its energy in the soft X-ray
band, the gas, tidally torqued earlier on, settles onto a disc. It is
subsequently contaminated by galactic winds ejected from the forming
galaxy but, according to the simulations of C10, the contamination is
small. The temperature of the gas is determined by the gravitational
potential of the halo and, for present day $L_\star$ galaxies, it is
of the order of $10^6\K$. Thus, a strong correlation is established
between coronal X-ray luminosity and halo mass. In the absence of
significant differences between the hot gas fractions of disc and
elliptical galaxies, this relation should not be sensitive to the
optical morphology of the galaxy.

In this study, we have obtained strong observational support for this
general picture. Firstly, we report a correlation between coronal X-ray
luminosity and stellar mass (as measured by $K$-band luminosity) that
has essentially the same normalisation, slope and scatter for disc and
elliptical galaxies.  Secondly, we report a correlation between coronal
X-ray luminosity and mass (as measured by the spectral temperature of
the plasma) that is also similar for both types of galaxy.

We stress that the correlations we have found involve the {\it
coronal} gas. They are not affected by point sources since the X-ray
emission has been explicitly corrected for non-thermal point source
contamination and, with a few exceptions, the X-ray luminosities are
much higher than the contribution expected from unresolved thermal
point sources (such as accreting white dwarfs and cataclysmic
variables) according to the relation derived by
\citet{Revnivtsev_et_al_08}. Thus, the correlations we have found are
qualitatively and quantitatively different from those previously
obtained between \textit{total} X-ray luminosity (including stellar
contributions) and optical properties as a function of galaxy
morphology \citep[e.g.][]{Fabbiano_89}.

While according to theory and simulations the coronal gas in both
discs and ellipticals is predominantly primordial, it is often thought
that the source of coronal gas could be internal, namely gas ejected
during stellar evolution and heated mostly by Type~II SNe, in the case
of discs, and by Type~Ia SNe, in the case of ellipticals
\citep[e.g.][]{Mathews_and_Brighenti_03,Tullmann_et_al_06}. In this
picture, it is difficult to understand why discs and ellipticals
should have such similar \lxlk\ and \lxTx\ relations given that they
have such different stellar populations and star formation rates. In
the Appendix, we present analytic arguments, supported by
observational evidence, that demonstrate the difficulty of
establishing common
\lxlk\ and \lxTx\ relations through such mechanism.

By contrast, in our picture in which the X-ray emission arises from
coronal gas, the correlations between $L_{\rm X}$, $T_{\rm X}$ and
$K$-band luminosity in present day $L_\star$ galaxies are easy to
understand: they arise because all three quantities are proportional
to halo mass. This is directly seen in the \gimic\ simulations
\citep[][C10]{Crain_et_al_09_short}, which also show how 
the scatter in the \lxlk\ relation arises from the scatter in the
\lk-\mvir\ relation (see Fig.~6 of C10).

We remarked in \S\,\ref{sec:LxTx} that the extended \lxTx\ relation,
obtained by combining our sample of galaxies with data for galaxy
groups and galaxy clusters
\citep{Helsdon_and_Ponman_00,Mulchaey_et_al_03,Horner_01}, can be
described by a broken power-law, with the break at $\sim1\keV$. A
steepening of the
\lxTx\ relation at galaxy group temperatures has been known for some
time \citep[e.g.][]{Helsdon_and_Ponman_00}, but we now see that the
steeper slope extends seamlessly down to normal galaxies\footnote{ The
break in the \lxTx\ relation apparent in Fig.~\ref{fig:LxTx} is not
due to the use of soft ($0.5-2.0\keV$), rather than bolometric, X-ray
luminosities. We have checked this by constructing a version of
Fig.~\ref{fig:LxTx} for bolometric luminosities estimated assuming a
bolometric correction for an \apec\ plasma model of the observed
temperature and metallicity.  The bolometric relation still shows a
prominent break at $\sim1\keV$, and both galaxies and galaxy groups
exhibit similarly steep relations.}. Extensive theoretical work,
extending back over a decade, has sought to explain the origin of the
steepening of the relation at group temperatures (or masses). Simple
preheating models, in which the entropy of the proto-intragroup and
proto-intracluster media is uniformly raised by some unspecified
feedback source, are able to reproduce the break
\citep[e.g.][]{Balogh_Babul_and_Patton_99}. 

A qualitatively similar relation to the one shown in
Fig.~\ref{fig:LxTx}, with a break at $0.7 \keV$, was obtained by
\citet{Dave_Katz_and_Weinberg_02} 
in a cosmological hydrodynamic simulation including radiative cooling
and star formation. These authors identified two causes for the break:
i)~a reduction in the X-ray luminous gas density in haloes below the
break scale due to the removal of coronal gas, and ii)~ a systematic
variation of the density structure of the corona with halo mass. As
discussed in \S\,\ref{sec:introduction}, a reduction in the central
gas density is one of the reasons why the gas coronae of $L_\star$
galaxies in the
\gimic\ simulations are far less luminous than expected
by WF91. On the scale of galaxies, however, this reduction is
primarily effected by feedback from Type~II SNe, rather than by
cooling and star-formation.

A common origin for the hot coronal gas in $L_\star$ disc and
elliptical galaxies raises an interesting question: if the X-ray
properties of the coronae are so similar, why are the star formation
properties of discs and ellipticals so different? Although we plan to
address this conundrum using simulations such as \gimic, it is
tempting to relate the differences to the processes that turn discs
into present day ellipticals. According to \citet{Parry_Eke_and_Frenk_09}, the
most common processes are minor mergers and disc instabilities (except
for bright ellipticals which form predominantly by mergers) occuring
at $z \lsim 1$. These events are likely to be accompanined by
starbursts that produce a relatively small number of stars but that
can temporarily disrupt the cooling flow from the corona. The flow is
eventually restored but only a relatively a small amount of gas has
had time to cool by the present. A mechanism of this sort might
explain the blue discs around bulge-dominated galaxies detected using
\galex\ data by \citet{Kauffmann_et_al_07_short} and the low-level star
formation inferred in most ellipticals by \citet{Kaviraj_et_al_08_short}. Searches
for cold gas with millimetre and radio wave facilities
\citep[e.g.][]{Oosterloo_et_al_07,
Combes_Young_and_Bureau_07,Krips_et_al_10} would provide a useful test
of this picture.

Our conclusions are based on the analysis of a small and heterogeneous
sample, particularly in the case of disc galaxies. Homogeneous samples
of \textit{optically} selected, normal disc and elliptical galaxies
can easily be extracted from surveys such as the SDSS and their X-ray
properties determined from deep X-ray observations that are possible
with \chandra\ and \xmm. Such a programme would set new and valuable
constraints on theories of galaxy formation.


\section*{Acknowledgements}  
\label{sec:acknowledgements}

We thank John Mulchaey for supplying us with unpublished temperature measurements. RAC acknowledges the hospitality of the Institute for Computational Cosmology, Durham, and the Institute of Astromomy, Cambridge, where part of this work was carried out. RAC is supported by the Australian Research Council through a Discovery Project grant. IGM acknowledges support from a Kavli Institute Fellowship at the University of Cambridge. CSF acknowledges a Royal Society Wolfson Research Merit award. This study makes use of data products from the Two Micron All Sky Survey (\twomass) and the Infrared Astronomical Satellite (\iras), obtained from archives hosted by \ipac.


\bibliographystyle{mn2e}
\bibliography{bibliography} 
\bsp


\begin{appendix}
\medskip

\section{Supernovae heating and hot gaseous coronae}
\label{sec:sne_heating}

We briefly investigate the role of internal SNe heating in the
production of hot circumgalactic gas around $L_\star$ galaxies. We
first consider star-forming disc galaxies in which Type~II SNe
dominate the energy injection rate.

There are some galaxies in which extra-planar X-ray emission is
clearly being powered by an outflow driven by Type~II SNe, for
example, where there is co-spatial optical line emission such
as H$\alpha$ (e.g., Str04). In the most dramatic examples such as M82,
biconical X-ray contours \citep{Strickland_and_Heckman_07} are the
tell-tale sign of strong nuclear outflows \citep[for a review,
see][]{Veilleux_Cecil_and_Bland-Hawthorn_05}. However, the fraction of
detectable extra-planar X-ray emission associated with outflows is
uncertain. 

C10 observed that the \lxsfr\ correlation is signficantly weakened
when disc galaxies with low X-ray luminosity and low star formation
rate are included. As shown in their Fig.~5, observations of
star-forming galaxies exhibit considerable scatter in the \lxsfr\
plane. If the hot gas associated with disc galaxies were exclusively
heated by Type II SNe, a stronger correlation between
\lx\ and \sfr\ would be expected, because of the prompt  detonation of
Type~II SNe after star-formation episodes ($\lesssim 30\Myr$) and the
relatively short central cooling time of the corona ($t_{\rm
cool}^{\rm corona} \ll t_{\rm H}$). Morever, as argued by C10, even a
strong correlation between \lx\ and \sfr\ does not necessarily imply
that the hot gas is related to SNe since (to first order) both
quantities are expected to scale with the mass of the galaxy's dark
matter halo.

In present-day ellipticals, with little or no ongoing star formation,
Type Ia SNe are the dominant energy source. Integrating over the
lifetime of the observed stellar populations, it would seem that the
energy budget from Type~Ia SNe is sufficient to maintain a
quasi-hydrostatic hot corona
\citep{Mathews_and_Baker_71,Loewenstein_and_Mathews_87,Brighenti_and_Mathews_99}.
In \S\,\ref{sec:interpretation}, however, we suggested that the
similarity of the \lxlk\ and \lxTx\ relations for disc and elliptical
galaxies argues against the hypothesis that the coronae of the two
types of galaxy are produced by different mechanisms. 

Let us set aside for the moment the argument that the scatter in the
\lxsfr\ relation of disc galaxies is difficult to reconcile with an
internal heating origin for hot coronae. For both morphological types
to produce the same X-ray luminosity at fixed stellar mass and fixed
halo mass, it is necessary that the energy injection rate at the
present day, from Type~II SNe in the case of disc galaxies and from
Type~Ia SNe in the case of ellipticals galaxies be comparable. This
requires uncomfortable fine-tuning.

Using empirical constraints, we can estimate whether such a
coincidence is possible. Adopting the \citet{Chabrier_03} stellar initial
mass function (IMF), spanning the range $0.1$-$100\Msun$, let us
assume that all stars with masses between $8$ and $100\Msun$ end their
lives as Type~II SNe, and that 2.5~percent of stars with masses between
$3$ and $8\Msun$ end their lives as Type~Ia SNe \citep[see Fig. A6
of][]{Wiersma_et_al_09}. Each SN, of either type, generates  a kinetic
energy, $E_{\rm SN}$, which we will assume to be $\sim10^{51}\erg$. In
the case of Type~II SNe, this energy is liberated on a timescale much
shorter than ${\rm min}(t_{\rm dyn},t_{\rm cool})$, and the rate of
energy injection from these events can thus be approximated as
\begin{equation}
 \dot{E}_{\rm SNII}(M_\star) = E_{\rm SN} \dot{M}_\star(M_\star)
\int_{8\Msun}^{100\Msun} \phi(m) {\rm d}m,  
\label{eq:TypeII}
\end{equation}
where $\phi(m) {\rm d}m$ is the IMF. A reasonable estimate of
$\dot{M}_\star(M_\star)$ can be obtained from a simple power-law fit
to the $\dot{M}_\star-M_\star$ plane of SDSS star-forming galaxies
derived by \citet[][their Fig.~17]{Brinchmann_et_al_04}. We therefore
adopt  
\begin{equation}
\log_{10}\dot{M}_\star [\Msunyr] = -5.865 + 0.615\log_{10} M_\star [\Msun].
\label{eq:Brinchmann04}
\end{equation}

Since Type~Ia SNe are thought to result from binary evolution, a
single stellar population will produce Type~Ia SNe over an extended
period. The lifetimes of the progenitors remain poorly understood
\citep[see e.g.][]{Podsiadlowski_et_al_08}, requiring that the number
of detonations per unit time be modelled with a theoretical or an
empirically-motivated delay function, $\xi(t)$, which is normalised
such that $\int_0^\infty\xi(t){\rm d}t=1$. Hence, the specific number
of SNIa over some time interval is,
\begin{equation}
n_{\rm SNIa}(t:t+\Delta t) = \nu \int_t^{t+\Delta t}\xi(t'){\rm d}t',
\label{eq:TypeIa_detonations}
\end{equation}
where $\nu$ is the number of Type~Ia SNe per unit stellar mass formed
that will ever occur. Recall that we adopt a value of 2.5 percent of
all stars between $3$ and $8\Msun$, such that
\begin{equation}
\nu = 0.025 \int_{3\Msun}^{8\Msun} \phi(m) {\rm d}m.
\label{eq:nu}
\end{equation}
We assume a simple e-folding delay function of the form,
\begin{equation}
\xi(t) = \frac{e^{-t/\tau}}{\tau},
\label{eq:e_folding}
\end{equation}
where $\tau$ is the characteristic delay time for which we take a
fiducial value of $3\Gyr$, which was shown by \citet[][see their
Fig.~A6]{Wiersma_et_al_09} to reproduce, in cosmological simulations,
the observed cosmic Type~Ia SN rate
\citep[e.g.][]{Cappellaro_Evans_and_Turatto_99,Hardin_et_al_00_short,Pain_et_al_02_short,Madgwick_Hewett_and_Mortlock_03,Tonry_et_al_03_short,Blanc_et_al_04_short,Dahlen_et_al_04_short,Barris_and_Tonry_06,Neill_et_al_06_short,Poznanski_et_al_07_short,Kuznetsova_et_al_08_short}.
For simplicity, we assume the spread of Type~Ia ages contributing at
any given time is $\ll \tau$. Hence
\begin{equation}
 \dot{E}_{\rm SNIa}(M_\star) = \nu E_{\rm SN}M_\star \int_{t_{\rm
age}}^{t_{\rm age}+{\rm d}t} \frac{e^{-t'/\tau}}{\tau}\frac{{\rm
d}t'}{{\rm d}t}, 
\label{eq:TypeIa}
\end{equation}
\begin{figure}
\includegraphics[width=\columnwidth]{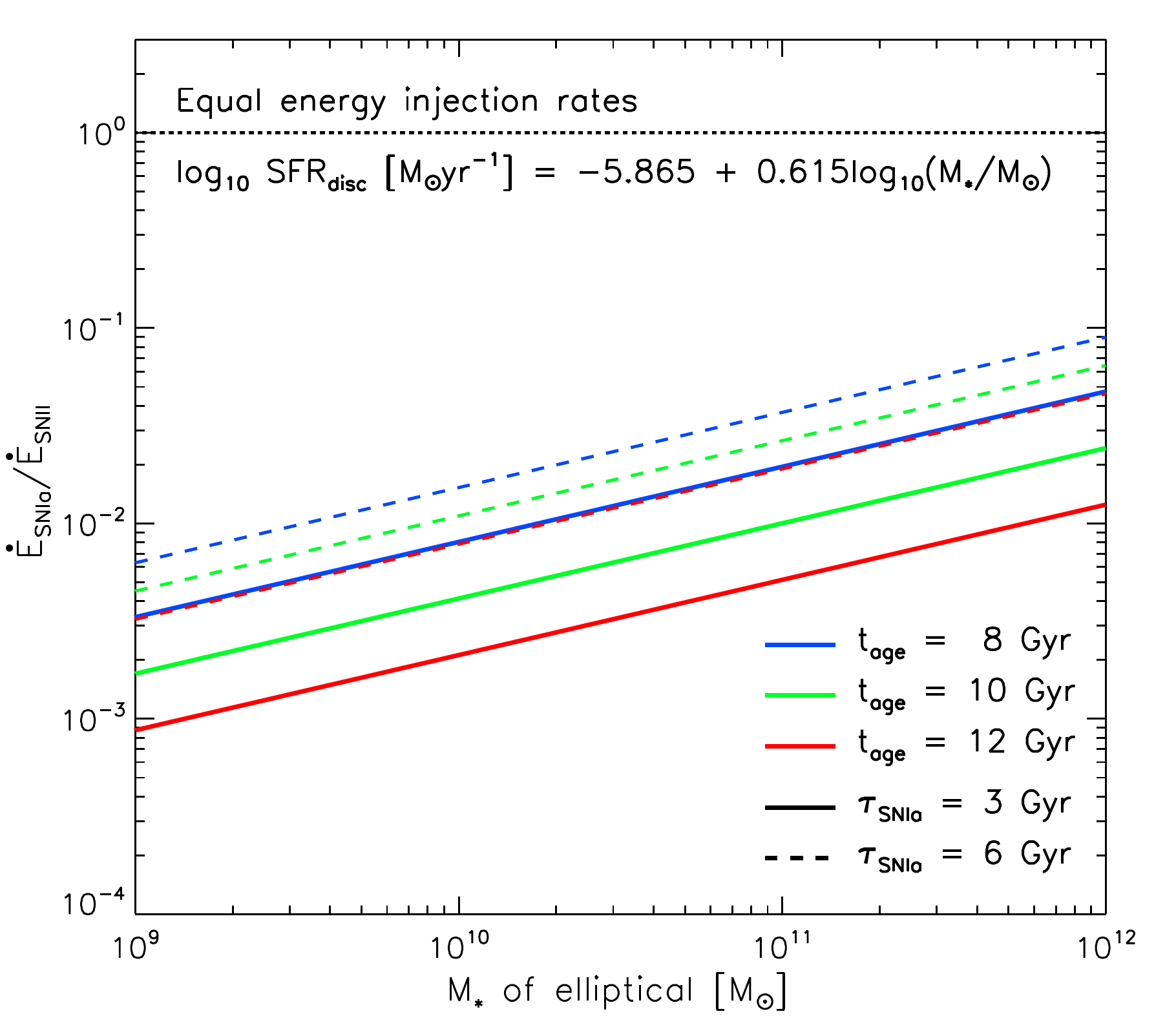}
\caption{The ratio of instantaneous energy injection rates from
Type~Ia SNe, derived from an evolved stellar population within an
elliptical galaxy, and Type~II SNe from ongoing star formation at the
average rate observed in SDSS galaxies
\citep{Brinchmann_et_al_04}. Different ages for the dominant stellar
population in the elliptical galaxy $(8,10,12\Gyr)$ are shown by the
coloured lines. The solid and dotted lines are for an assumed
characteristic delay time for Type~Ia SNe of $3\Gyr$ and $6\Gyr$
respectively.}
\label{fig:SNe_energy_ratios}
\end{figure}
Within observational constraints, we have some flexibility in the
assumed age of the dominant stellar population in an elliptical
galaxy, $t_{\rm age}$, so in Fig.~\ref{fig:SNe_energy_ratios} we adopt
three values ($8$, $10$ and $12\Gyr$) and plot the ratio of the energy
injection rates from Type~Ia SNe from these populations to that of
Type~II SNe (Eqn.~\ref{eq:TypeII}), as a function of the galaxy's
stellar mass. The solid coloured lines show that, for reasonable
choices of the age of the elliptical galaxy population and our
well-motivated (but uncertain) choice of the e-folding timescale
($3~\Gyr$), the energy injection rate due to Type~II SNe is always
$\sim 1-2$ orders of magnitude greater than that due to the Type~Ia
SNe of an evolved population.

The only reasonable freedom we have to vary the parameters of this
simple model is to modify the poorly constrained value of $\tau$. A
longer delay timescale shifts a greater fraction of the energy
liberated by Type~Ia SNe to later times so, for reasonable elliptical
galaxy formation histories, a greater value of $\tau$ will increase
$\dot{E}_{\rm SNIa}$ at $z=0$. We therefore also consider the
bracketing case of doubling the delay timescale (to $6~\Gyr$), shown
in Fig.~\ref{fig:SNe_energy_ratios} with dashed lines. Clearly, making
this conservative assumption does not alter our main result. We
therefore conclude that it is not possible to accomodate a model in
which the similar X-ray luminosities of normal disc-dominated and
elliptical galaxies at fixed stellar mass stem from the ongoing
injection of energy from these two different internal sources.

\end{appendix}

\label{lastpage}
\end{document}